# Development of an acoustic transceiver for the KM3NeT positioning system

G. Larosa, M. Ardid, C.D. Llorens, M. Bou-Cabo[*], J.A. Martínez-Mora, S. Adrián-Martínez[*] (for the KM3NeT Consortium)

*Institut d'Investigació per a la Gestió Integrada de Zones Costaneres (IGIC)-Universitat Politècnica de València, C/Paranimf 1, 46730 Gandia, València, Spain*
[*] Multidark fellow

**Abstract**

In this paper we describe an acoustic transceiver developed for the KM3NeT positioning system. The acoustic transceiver is composed of a commercial free flooded transducer, which works mainly in the 20-40 kHz frequency range and withstands high pressures (up to 500 bars). A sound emission board was developed that is adapted to the characteristics of the transducer and meets all requirements: low power consumption, high intensity of emission, low intrinsic noise, arbitrary signals for emission and the capacity of acquiring the receiving signals with very good timing precision.

The results of the different tests made with the transceiver in the laboratory and shallow sea water are described, as well as, the activities for its integration in the Instrumentation Line of the ANTARES neutrino telescope and in a NEMO tower for the in situ tests.

## 1. Introduction

KM3NeT is a European Consortium with the goal to build and operate a multi-cubic-kilometre detector in the Mediterranean Sea. It will be composed from hundreds semi-rigid structures, containing optical modules, anchored on the seabed and maintained vertical with a buoy [1]. An Acoustic Positioning System (APS) is necessary to monitor the positions of all optical modules in the deep sea, as marine currents cause inclinations of the structures. Thus, the optical modules can be displaced up to several meters from their nominal positions. Precise knowledge of the relative positions of the optical modules (precision of about 10 cm) is needed for an accurate reconstruction of the muon tracks produced in neutrino interactions with the matter around the detector. A big effort has been made during the last years to develop an APS to the future KM3NeT neutrino telescope [4-6]. With respect to other commercial systems (as the ones used in ANTARES and NEMO) the new developments aim for a better integration to the KM3NeT infrastructure, cost reduction, more profound control of the system, better tuning of the acoustic power, and the possibility to use the acoustic data for other studies. The detector will have dimensions larger than 1 km$^3$, so the acoustic transceiver presented in this paper has been developed to attend to the distances involved (1−2 km), and to constraints imposed by a deep-sea neutrino telescopes (hostile environmental conditions: high pressure, corrosion, etc.; technology standards: low-power consumption, specific communication protocols, etc.). Moreover, a transceiver prototype has been integrated in the Instrumentation Line of the ANTARES neutrino telescope [2] to be tested in situ. It is also being integrated in NEMO phase II tower [3]. These tests are very important for the final implementation in the KM3NeT detector.

## 2. Acoustic transceiver

The acoustic transceiver is constituted of a transducer and an electronic board named S*ound Emission Board* (SEB), shown in Fig. 1. The transducer type is a Free Flooded Ring (FFR), model SX30 manufactured by Sensor Technology Ltd. It withstands high pressure (tested up to 440 bars) [7-8]. The FFR has dimensions of 2.5 cm height, 4.4 cm and 2 cm outer and inner diameter respectively. The operating frequency range is 20-40 kHz with good receiving sensitivity and transmitting power (-193 dB re 1 V/µPa and 133 dB re 1 µPa/V at 1 m, respectively). The beam pattern is omnidirectional in the plane perpendicular to the axis of the ring, while in the planes containing the axis there is a minimum (reduction of 5 dB) of sensitivity responses at 60º. Moreover, the electronic noise is about -130 dB re 1 V/(Hz)$^{1/2}$ with a maximum input power of 300 W with a 2% duty cycle.

The SEB [9] has been designed to match the characteristics of the FFR and for the requirements of the KM3NeT detector. The diagram of the SEB, which consists of three parts, is shown in Fig. 1. The first part, bottom part of the diagram, is a communication and control block which contains the microcontroller dsPIC[1]. It manages the emission and the reception tasks and the settings, for instance setting the board with arbitrary waveforms. The emission block, in the middle, consists of the digital amplification plus the transducer impedance matching. It is able to store 1.6 J of energy for emission. Whit the energy stored on the capacitor it is possible to emit signals of 3 ms length (at the maximum power) having a voltage drop smaller than 1 V. Successively the transformer converts the input digital signal to an output signal with higher voltage (in the range 35–400 $V_{pp}$) in order to have enough amplitude to feed the transducer for the emission of the acoustics signals enabling to compensate for attenuation over the large distances involved in the KM3NeT detector. The last part, shown on top, is the reception part. The main component is a relay to use the transducer as receiver. Furthermore, the designed SEB allows the transducer to emit high amplitude short signals (a few ms length) with arbitrary waveform. Moreover, it has been designed for low-power consumption and adapted to the deep-sea neutrino telescope infrastructure using power supplies of 12 V and 5 V with a consumption of 1 mA and 100 mA respectively. To avoid initial high currents, there is a current limit of 15 mA when the capacitor starts to charge from the 12 V line. Few seconds later the current stabilizes at 1 mA. The communication of the board with a control PC is established through the standard RS232 protocol. In order to have very good timing synchronization the emission is triggered using a LVDS bus controlled by the dsPIC with stability better than 1 µs [8].

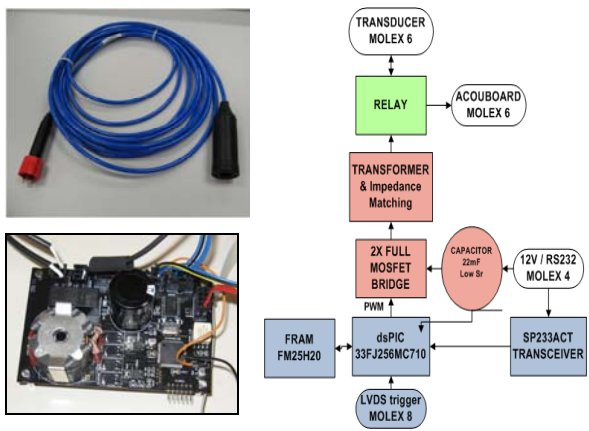

Figure 1: Views of the Free Flooded Ring transducer, of the Sound Emission Board, and its diagram.

## 3. Tests of the transceiver

Different tests have been performed with the acoustic transceiver in order to check that the specifications are met. For instance, pressure tests were performed and reported in [8]. Other tests were performed in a tank of 87.5 x 113 x 56.5 $cm^3$, and in a pool of 3.6 x 6.3 x 1.5 $m^3$ (fresh water in both cases) to study the transmitting voltage response and the directional response. The results have been reported in [5]. Here, we report the tests performed in the Gandia Harbour (Spain) to check the performance of the transceiver over longer distances and the response in a noisy environment. Particularly, we were interested in studying the use of different acoustic signals for positioning purposes. We were confident that the use of wideband signals, Maximum Length Sequence (MLS) signals and sine sweep signals, instead of pure sinusoidal signals may result in an improvement of the signal-to-noise ratio, and therefore resulting in an increase in the detection efficiency, as well as in the accuracy of the time of detection.

The transceiver (SX30 FFR plus SEB) was used as emitter. A second SX30 FFR with a RESON[2] preamplifier, model CCA1000 (20 dB gain), was used as receiver hydrophone. Here, we show the results for a distance of about 140 m between the emitter and receiver. Figure 2 shows the receiving signals using a MLS signal (order: 11, sampling frequency: 200 kHz), a linear sine sweep signal (frequency range: 20-48 kHz, length: 4 ms), and a pure sinusoidal signal (frequency: 30 kHz, length: 4 ms). The signals were recorded using an external trigger synchronization system between the emitter and the receiver. The system produces a delay of 8.25±0.03 ms in the receiver channel. The direct signal arrives at about 0.085 s. Since the noise is quite high (~Pa), and there are reflections, the signals are not easy to identify. However, as shown in Figure 3, looking at the correlation between the receiving and the emitted signal, the time when the signal appears is clearly observed. For the case of the MLS and sine sweep signals, a clear thin peak is observed, and therefore it is easy to determine the time of detection. The other peaks are due either to reflections or to the noise. The measurements were performed three times obtaining a time of detection of 85.08±0.03 ms (~4.5 cm uncertainty) for the MLS signal and of 85.075±0.015 ms (~2.3 cm uncertainty) for the sine sweep signal. Notice that the accuracy is similar to that of the synchronization system. Therefore, these measurements do not allow for the determination of the time detection uncertainty of the APS prototype, but the measurements are compatible with those taken previously in the lab with uncertainties on the order of a microsecond. The case of the pure sinusoidal signal is completely different. As shown in Figure 3, a very broad peak is obtained, the time of the maximum being quite sensitive to noise or reflections. Following the previous approach, an uncertainty of the millisecond order is obtained, and therefore this method is not the best one for this kind of signals. Usually, a band-pass frequency filter is applied, and the detection time is determined by reaching a threshold level [11-12]. Doing this properly requires a very accurate calibration in order to determine the inertial delay of the hydrophone, and even so, it can give bad results in case of high noise or intense reflections nearby that can add to the waves constructively, as it happened during our measurement in the harbour.

---

[1] http://www.microchip.com

[2] http://www.reson.com

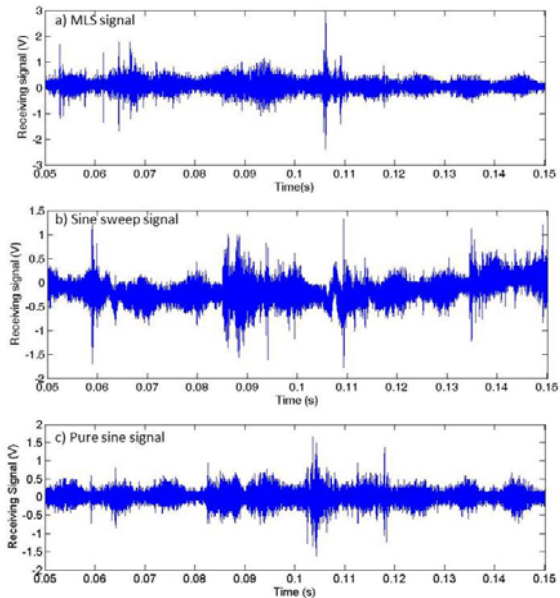

Figure 2: Receiving signal using three different kinds of signals.

In contrast, the cross-correlation of broadband signals is less sensitive to these effects. The inertial delay, which affects mainly to the start and end of the signal, is rendered less important by considering the whole duration of the signal. The effect of the reflections is reduced by distinguishing between different peaks of the cross-correlated signal, the first main peak being the one to consider.

Finally, in order to perform an in situ test, the system has been integrated in the active anchor of the Instrumentation Line of the ANTARES detector. The SEB was installed in a container which also houses a Laser system used for timing calibration purposes. A new functionality for the microcontroller of the SEB was implemented to control the laser emission as well The FFR hydrophone was fixed in the base of the line at 50 cm from the standard emitter transducer of the ANTARES positioning system with the area opposite to the moulding and the cable of the hydrophone looking upwards. It has been fixed through a support of polyethylene. The Instrumentation Line was successfully deployed at 2475 m depth on 7th June 2011 at the nominal target position. However, the connection of the Line to the Junction Box has been delayed and will be done when the ROV[3] will be available (probably in April 2012). Once the line will be connected, the transceiver can be fully tested in real conditions. In addition, in April 2012 another transceiver prototype will be integrated in the NEMO phase II tower. As in the previous case, it will be fixed in the tower base and the SEB will be located inside a titanium container holding other electronic parts and a laser.

## 4. Summary and Conclusions

An APS is needed in a deep-sea neutrino telescope and we have presented here studies and improvements done to develop a transceiver that will be implemented in KM3NeT. The tests and measurements done with the transceiver in the Gandia harbour do not allow for a measurement of the time detection uncertainty of the APS prototype with good precision, but allows us to say that the accuracy is better than 30 μs and the measurements are compatible with those taken previously in the lab with uncertainties on the order of a microsecond. Then, we can conclude that the transceiver seems to satisfy the requirements and accuracy needed for the APS. Moreover, the transceiver can handle a transmitting power above 170 dB re 1μPa@1m. Combined with adequate signal processing techniques the APS may cover the large distances involved in a neutrino telescope.

The system will be integrated in NEMO phase II tower in 2012 and it has been integrated in the ANTARES neutrino telescope (waiting for the connection of the Instrumentation Line) to test the transceiver in situ in the deep-sea.

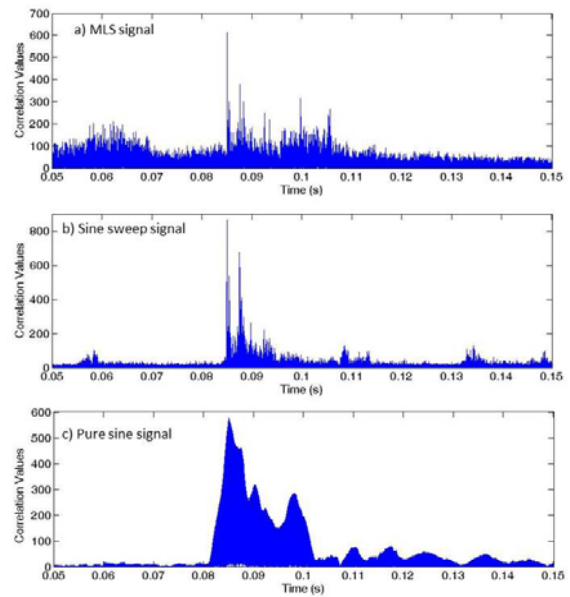

Figure 3: Correlation between emitted and received signals.

**Acknowledgments**

This work has been supported by the Ministerio de Ciencia e Innovación (Spanish Government), project references FPA2009-13983-C02-02, ACI2009-1067, AIC10-D-00583, and Consolider-Ingenio Multidark (CSD2009-00064). It has also been funded by Generalitat Valenciana, Prometeo/2009/26, and the European 7th Framework Programme, grant no. 212525.

---

[3] ROV: Remotely Operated Vehicle.